\documentclass{aa}
\usepackage{graphicx}
\usepackage{hyperref}
\usepackage{txfonts}
\begin{document}

   \title{Hemispheric sunspot numbers 
   1874--2020}


   \author{Astrid M. Veronig\inst{1,2}
   \and Shantanu Jain\inst{3}
   \and Tatiana Podladchikova\inst{3}
   \and Werner P\"otzi\inst{2}
   \and Frederic Clette\inst{4}
      }

\institute{University of Graz, Institute of Physics, Universit\"atsplatz 5, 8010 Graz, Austria\\ 
\email{astrid.veronig@uni-graz.at}
\and
University of Graz, Kanzelh\"ohe Observatory for Solar and Environmental Research, Kanzelh\"ohe 19, 9521 Treffen, Austria
\and
Skolkovo Institute of Science and Technology, Bolshoy Boulevard 30, bld. 1, Moscow 121205, Russia
\and
World Data Center SILSO, Royal Observatory of Belgium, 1180 Brussels, Belgium 
}

   \date{Received Month , Year; accepted Month, year}

  \abstract
   {Previous studies show significant north--south asymmetries for various features and indicators of solar activity. These findings suggest some decoupling between the two hemispheres over the solar cycle evolution, which is in agreement with dynamo theories. For the most important solar activity index, the sunspot numbers, so far only limited data are available for the two hemispheres independently.    }
   {The aim of this study is to create a continuous series of daily and monthly hemispheric sunspot numbers (HSNs) from 1874 to 2020, which will be continuously expanded in the future with the HSNs provided by SILSO.}
   {Based on the available daily measurements of hemispheric sunspot areas from 1874 to 2016 from Greenwich Royal Observatory and 
   National Oceanic and Atmospheric Administration (NOAA), we derive the relative fractions of the northern and southern activity. These fractions are applied to the international sunspot number (ISN)
   to derive the HSNs. This method and obtained data are validated against published HSNs for the period 1945--2004 and those provided by SILSO for 1992 to 2020.}  
   {We provide a continuous data series and catalogue of daily, monthly mean, and 13-month smoothed monthly mean HSNs for the time range 1874--2020 ---fully covering solar cycles\ 12 to 24--- that are consistent with the newly calibrated ISN \citep{clette2014}. Validation of the reconstructed HSNs against the direct data available since 1945 reveals a high level of consistency, with Pearson correlation coefficients of $r=0.94$ (0.97) for the daily (monthly mean) data.
   The cumulative hemispheric asymmetries for cycles 12--24 give a mean value of 16\%, with no obvious pattern in north--south predominance over the cycle evolution.  The strongest asymmetry occurs for cycle no.\ 19, in which the northern hemisphere shows a cumulated predominance of 42\%. The phase shift between the peaks of solar activity in the two hemispheres may be up to 28 months, with a mean absolute value over cycles 12--24 of 16.4 months. The phase shifts reveal an overall asymmetry of the northern hemisphere reaching its cycle maximum earlier (in 10 out of 13 cases), with a mean signed phase shift of $-7.6$ months. Relating the  ISN and HSN peak growth rates during the cycle rise phase with the cycle amplitude reveals higher correlations when considering the two hemispheres individually, with $r \approx 0.9$. }
   {Our findings provide further evidence that to some degree 
   the solar cycle evolves independently in the two hemispheres, and demonstrate that empirical solar cycle prediction methods can be improved by investigating the solar cycle dynamics in terms of the HSN evolution.}

   \keywords{Sun  --
                sunspots  --
                solar activity
               }

   \maketitle
%

\section{Introduction}

The sunspot number is a measure of the global solar activity derived from visual observations of the Sun's visible disk. It provides the longest record of the evolution of solar activity, as it extends back to the very first independent, detailed observations of sunspots made four centuries ago by Galileo Galileo in Italy, Christoph Scheiner in Germany, Johann Fabricius in the Netherlands, and Thomas Harriot in England, very soon after the invention of the telescope in 1609. 

The sunspot number is based on the so-called Wolf number, calculated for a single observation as:
\begin{equation}
W= 10\, N_g + N_s
,\end{equation}
where $N_s$ is the total number of spots and $N_g$ the total number of sunspot groups on the solar disk  \citep{wolf1856}. 
However, the raw Wolf number from one observer will depend on the telescope, the personal experience and visual acuity, and also on randomly varying sky conditions. In order to derive a global index independent from individual observers, the international sunspot number (ISN) is calculated from the Wolf numbers collected from many observers on a daily basis (currently, a network of 80 stations worldwide).
This processing, described in detail in \cite{Clette2007}, is carried out by World Data Center Sunspot Index and Long-term Solar Observations (WDC-SILSO), which provides daily, monthly mean, and yearly mean ISN time-series dating back to the year 1700. These time-series provide a unique long-term record of the 11-year solar cycle, which supports a wide range of scientific applications, including many solar-cycle prediction methods \cite[e.g. review by][]{petrovay2020}. 

While the base ISN series provides a total number for the whole Sun, in January 1992, the WDC-SILSO  also started producing hemispheric sunspot numbers (HSNs) using the same statistical processing as for the ISN series. However, the HSNs are  based on a subset of about 50 SILSO stations that also record hemispheric counts in addition to the base total counts.
Those hemispheric numbers allow the solar cycle evolution to be traced  in the northern and southern hemispheres independently. In the framework of the first end-to-end re-calibration of the ISN time-series \citep{clette2014,Clette2016}, a new version of the ISN was released in July 2015 (Version 2.0). Simultaneously, the hemispheric numbers were also recalculated in accordance with the corrections determined for the total ISN.

More than one century ago, early studies comparing the activity of the northern and southern hemispheres already revealed significant asymmetries
\cite[see, e.g. review by][]{hathaway2015}. For instance, \cite{spoerer1889} and \cite{maunder1904} noticed that sunspots can be preferentially located in one of the hemispheres over extended periods.
 \cite{waldmeier1971} noted in his comprehensive observations that for the years 1959--1969 (cycles 19 and 20), a variety of solar activity features like sunspots, faculae, and prominences were much more numerous in the northern hemisphere than in the southern  hemisphere, and that the coronal brightness was  also distinctly higher in the northern hemisphere. 

Since then, hemispheric asymmetries of solar activity have been studied in detail in a variety of observations and activity indices, such as sunspot groups and areas \citep[e.g.][]{newton1955,carbonell1993,berdyugina2003,norton2010,McIntosh2013,deng2016}, sunspot numbers \cite[e.g.][]{temmer2002,temmer2006,Zolotova2010,chowdhury2019}, magnetic field distributions \cite[e.g.][]{antonucci1990,knaack2005}, flare occurrence \cite[e.g.][]{garcia1990,temmer2001,joshi2004,roy2020}, and so on. In general, these hemispheric asymmetries do not exceed 20\% \citep{norton2014}, but during extreme minima, like the Maunder minimum, they may attain very high values. In addition to the findings that the solar cycle maxima may have different strength and may be reached at different times in the two hemispheres, it has also been shown that the solar-cycle-related polar field reversals occur at different times in the northern and southern hemispheres \citep{durrant2003,Svalgaard2013}. The physical origin of these north--south asymmetries and phase shifts ---indicative of a partial decoupling of the two hemispheres--- is presumed to lie in the solar dynamo  \cite[e.g.][]{sokoloff1994,ossendrijver1996,norton2014,schuessler2018}. Therefore,  studies of the long-term evolution of the solar cycles  in the two separate hemispheres provide important information for the dynamo process underlying this evolution.
 
In this paper, we reconstruct the HSNs back to the year 1874 based on observations of sunspot areas and merge those with  existing HSN series available since 1945,  providing full coverage of cycles 12 to 24. The method is cross-validated against the HSNs for the overlapping time range (1945--2016), and the data are compiled into a catalogue of daily and monthly HSNs from 1874--2020. Finally, we demonstrate that the separate consideration of the dynamics of the sunspot numbers for the individual hemispheres provides  more distinct relations between  the cycle growth rate and its amplitude, which is relevant for empirical solar cycle prediction methods  \citep{cameron2008}.

\section{Data}

In this study, we create a continuous series and catalogue of daily,  monthly mean, and 13-month smoothed monthly mean HSNs back to the year 1874 based on three different data sets of HSNs and hemispheric sunspot areas as described in the following. A few direct records of HSNs from individual stations exist from solar cycle 18 onwards. For the period from January 1945 to December 2004.
\cite{temmer2006} derived HSNs from the daily sunspot drawings obtained at Kanzelh\"ohe Observatory \citep[KSO, Austria;][]{poetzi2016}\footnote{Sunspot drawing from Kanzelh\"ohe Observatory are accessible as scans at \url{http://cesar.kso.ac.at/synoptic/draw_years.php}} and Skalnate Pleso Observatory (Slovakia), and merged them to a common homogeneous data series.\footnote{The data catalogue from \cite{temmer2006} is available at \url{http://cdsarc.unistra.fr/viz-bin/cat/J/A+A/447/735}}  The data coverage of  daily observations from these two stations is 84\% \cite[see also, ][]{Rybak2004HvaOB}.

As mentioned in Sect.1, since 1992, in addition to the ISN, WDC-SILSO also collects HSNs from its worldwide network. 
In this case, the daily coverage is
100\%, as data from 20 to 40 stations are typically available each day, out of about 80 active stations.\footnote{Data access at: \url{http://sidc.be/silso/}, ``Data'' section} Moreover, those hemispheric numbers are directly attached to the long-term calibration of the ISN series, which extends back over multiple centuries. Therefore, this series provides the base reference for the backward reconstruction of the HSNs presented here.

Next to the visual sunspot numbers, another sunspot-based measure of solar activity, the total sunspot area, is available over the period May 1874 to September 2016. This alternate series is primarily based on photographic plates of the Sun produced between 1874 and 1976 by a small network of stations run by the Royal Greenwich Observatory (RGO) and using the same equipment and procedures. This collection by RGO ended in 1976, and was extended using sunspot areas from the US National Oceanic and Atmospheric Administration (NOAA), which are based on sunspot drawings from the SOON stations (Solar Observing Optical Network) run by the US Air Force. It is thus a composite series with two parts,  each spanning multiple decades. Text files containing the daily sunspot areas  (in units of millionths of a hemisphere, $\mu hem$) and monthly mean values are available for the full sun as well as separately for the northern and southern hemispheres.\footnote{\url{https://solarscience.msfc.nasa.gov/greenwch.shtml}}

\section{Hemispheric sunspot number reconstruction and validation}

\subsection{Reconstruction}

In 2015, the ISN was entirely recalibrated, and during this major change, an obsolete 0.6 down-scaling convention inherited from the 19th century was also eliminated (change of unit), which brings the new series to the same scale as modern observations \citep{Clette2016}. The current version of the SILSO HSN (2.0) is consistent with this re-calibration. The corrections are variable in time, but overall, over the period for which those hemispheric numbers are available, i.e.\ from 1992 onward, the average scale change between versions 1 and 2 is around 1.44.

The HSNs compiled in \citet[][thereafter T06]{temmer2006} were derived by calculating the relative fractions of the sunspot numbers for the northern and southern hemispheres from the sunspot drawings at Kanzelh\"ohe and Skalnate Pleso Observatories, 
and normalising to the ISN ---which at that time was still the original series, now numbered version 1--- before recalibration. Therefore, in order to derive a continuous long-term series, we now need to renormalise them to the current recalibrated ISN (version 2.0), which has become the new official reference since 2015.
The renormalisation is done by deriving the relative fractions $n_i$ and $s_i$ of the daily sunspot number in the northern $(N_{\rm i,V1})$ and southern $(S_{\rm i,V1})$ hemisphere, respectively, from the T06 data set (which were calibrated at the time of their creation to ISN version V1.0) as
\begin{equation}
n_i = \frac{N_{\rm i,V1}}{S_{\rm i,V1} + N_{\rm i,V1}}
\label{eq:2}
,\end{equation}
and
\begin{equation}
s_i = \frac{S_{\rm i,V1}}{S_{\rm i,V1} + N_{\rm i,V1}}
\label{eq:3}
,\end{equation}
and applying these fractions to the daily recalibrated (version V2.0) ISN $(R_i)$ from SILSO to obtain the recalibrated north $(R_{N,i})$ and south $(R_{S,i}$) HSN as
\begin{equation}
R_{N,i} = n_i\cdot R_{i} 
\label{eq:4}
,\end{equation}
and
\begin{equation}
R_{S,i} = s_i\cdot R_{i} 
\label{eq:5}
,\end{equation}
so that  $R_{N,i}+R_{S,i}=R_i$.
With this procedure we have established a continuous series of HSNs that are consistent with the recalibrated ISN, from 1/1945 to date, which is purely based on records of the HSN. 

To reconstruct the HSNs further back in time, we make use of the data for the total and hemispheric sunspot area that are available for the time range May 1874 to September 2016 from RGO and NOAA. Our rationale is the same as for the renormalisation of the HSNs to the recalibrated sunspot numbers. We derive the relative fraction of the sunspot areas observed in the northern and southern hemispheres (analogously to Eqs. \ref{eq:2}--\ref{eq:3}), and apply these fractions to the ISN (Eqs. \ref{eq:4}--\ref{eq:5}). 

\begin{figure}
 \centering
\includegraphics[width=0.99\columnwidth]{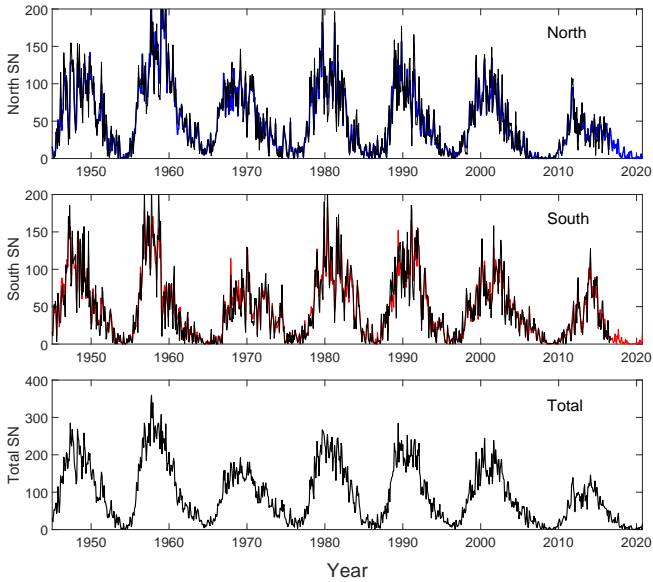}
\caption{From top to bottom: Monthly mean values of hemispheric north, south, and total sunspot numbers from 1945 to 2020. Top and middle panel: Blue (red) curves show the north (south) HSN series combined from T06 and SILSO, whereas the black curves show the HSNs reconstructed from the sunspot areas (available up to September 2016).
}
\label{fig:fig1}
\end{figure}

\begin{figure}
 \centering
\includegraphics[width=0.99\columnwidth]{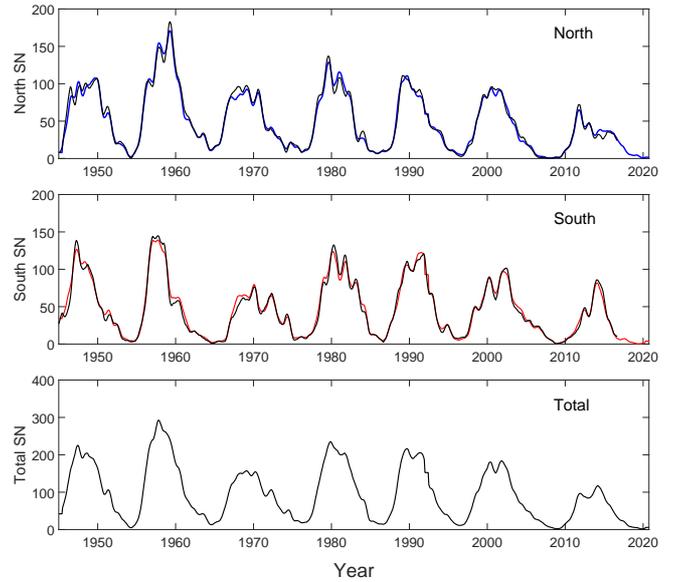}
\caption{Same as Fig. \ref{fig:fig1} but for the 13-month smoothed data. 
}
\label{fig:fig2}
\end{figure}

\subsection{Validation}

To validate this approach,
we evaluate for the overlapping period January 1945 to September 2016 how well the HSNs reconstructed from the sunspot area data fit with the hemispheric sunspot number series combined from T06 and SILSO. We note that for the combined hemispheric sunspot number series, we use the T06 data for the time range January 1945 to September 1991, and the HSNs from SILSO from January 1992 onward, because these latter
are based on a larger network of observatories and are regularly updated, allowing continuous prolongation of the hemispheric sunspot number data series established here.


Figure \ref{fig:fig1} (top panels) shows the HSNs reconstructed from the sunspot area records along with the T06 and SILSO HSNs for the time range 1945--2020. The bottom panel shows for comparison the total ISN.  Figure \ref{fig:fig2} shows the same for the 13-month smoothed monthly data. We note that we do not use the classical running mean but the smoothing method from \cite{podladchikova2017}, which provides an optimisation between the smoothness of the fitted curve and its closeness to the data (see also its application to the F10.7 and F30 radio flux data series in \cite{Petrova2021}). As can be seen from these plots, the overall correspondence between the original HSNs and the reconstructed data is high.

\begin{figure*}
\sidecaption
  \includegraphics[width=12cm]{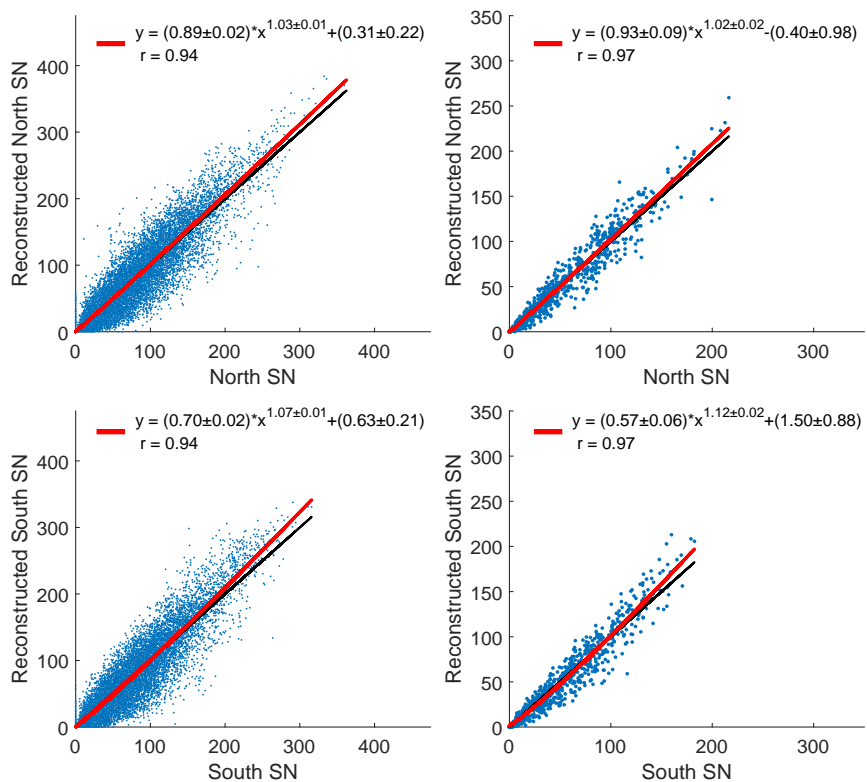}
     \caption{Scatter plots of daily (left) and monthly mean (right) HSNs reconstructed from sunspot areas against the HSNs from T06 and SILSO. Top and bottom panels show the northern and southern hemispheres.  The black line shows the ideal 1:1 correspondence and the red line represents the power-law fit. The inset gives the fit parameters and the Pearson correlation coefficient $r$.}
\label{fig:corr}
\end{figure*}

However, we note that the comparison of the monthly mean data suggests that on short timescales the HSN reconstructed from sunspot areas (Figure \ref{fig:fig1}) varies more than that provided by the SILSO and T06 data. Quantifying the fluctuations 
by deriving the point-wise differences of the monthly mean data for the two data sets, we find an rms of 27.2 (27.8) for the reconstructed north (south) HSN, whereas for the SILSO-T06 series the rms gives lower values of 19.4 (18.9).
There are two main effects that can contribute to this difference, and both are related to the properties of the sunspot area series which is contributing to our HSN reconstructions. The first one is related to 
differences in the measurement noise and the second is related to differences in the characteristic dynamics of the
sunspot number and area time-series:

a) Sunspot numbers are a smoother series than the sunspot areas because sunspot numbers are averaged over observations from more stations than the sunspot areas.
Therefore, the sunspot areas would have a larger `measurement noise' resulting in larger fluctuations. A discussion of the different noise contributions to the sunspot number series is given in \cite{Dudok2016}. In addition, in reconstructing the HSNs we add the errors from the two data series, that is, the ISN and the hemispheric sunspot areas. 

b) In the sunspot number series, all sunspots are counted as 1 and all groups as 10.  Therefore, the actual size or area of sunspots is not directly reflected in the sunspot number. It is only in the sense that, on average, the area increases with the number of sunspots that both measures follow  the same overall variations quite closely. In particular, in the sunspot number, small sunspots contribute a significant fraction to the total count, whereas in the total sunspot areas, the largest sunspots strongly dominate (while small sunspots play only a minor role). Large sunspots are typically long-lived, lasting several  weeks. On the other hand, small sunspots are short-lived and appear and disappear randomly from one day to the next. Therefore, the total area will tend to evolve more progressively than the sunspot number, which may explain the somewhat larger amplitudes and variations in the reconstructed HSNs.

The correspondence between the data series is further quantified in Figure  \ref{fig:corr}, where we show the scatter plots of the hemispheric sunspot data reconstructed from the sunspot area measurements against the corresponding values from the direct HSNs (combined T06 and SILSO data set) for the daily (left panels) and monthly mean (right panels) data for the time range January 1945 to September 2016. As one can see, the correlation coefficients are  high, with $r=0.94$ for the daily and $r=0.97$ for the monthly data.
The black line shows the 1:1 correspondence between both data sets, and the red line shows the result of a power-law fit. The power-law exponents derived range from 1.02 to 1.12, indicating some small non-linearity. As one can see from the scatter plots in Fig. \ref{fig:corr},  deviations from the linear 1:1 relation show up in the case of very high activity (large-amplitude cycles), with the reconstructed sunspot numbers revealing some trend to lie above the 1:1 line. As discussed below, this is most likely a residual effect of the non-linear relationship between the sunspot areas and sunspot numbers.

\begin{figure*}
 \centering
\includegraphics[width=0.99\textwidth]{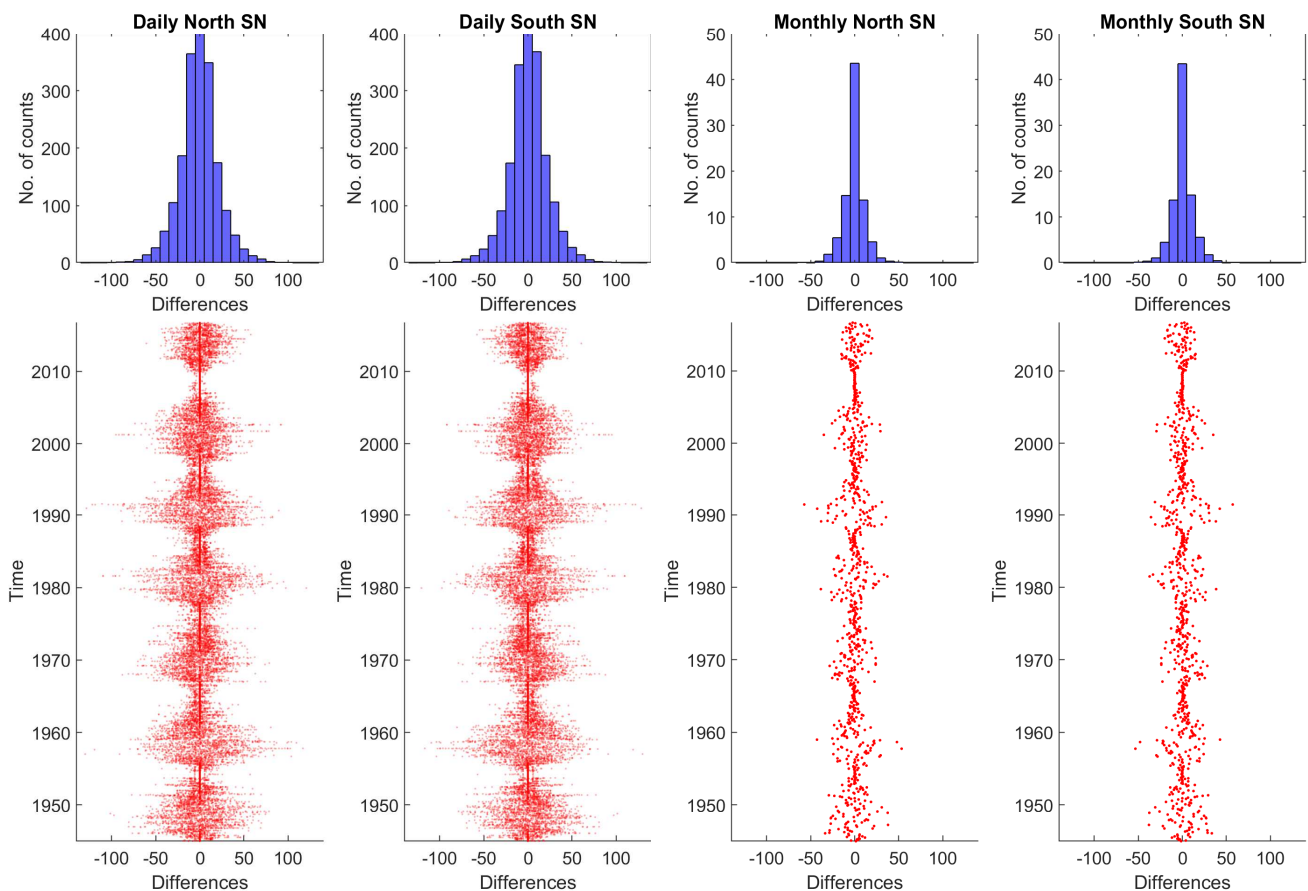}
\caption{Top: Histograms of the differences between the daily (left) and monthly (right) HSNs reconstructed from sunspot areas and the HSNs from T06 and SILSO. Bottom: Time evolution of the corresponding daily (monthly) differences.
}
\label{fig:fig4}
\end{figure*}

\begin{figure}
 \centering
\includegraphics[width=1.0\columnwidth]{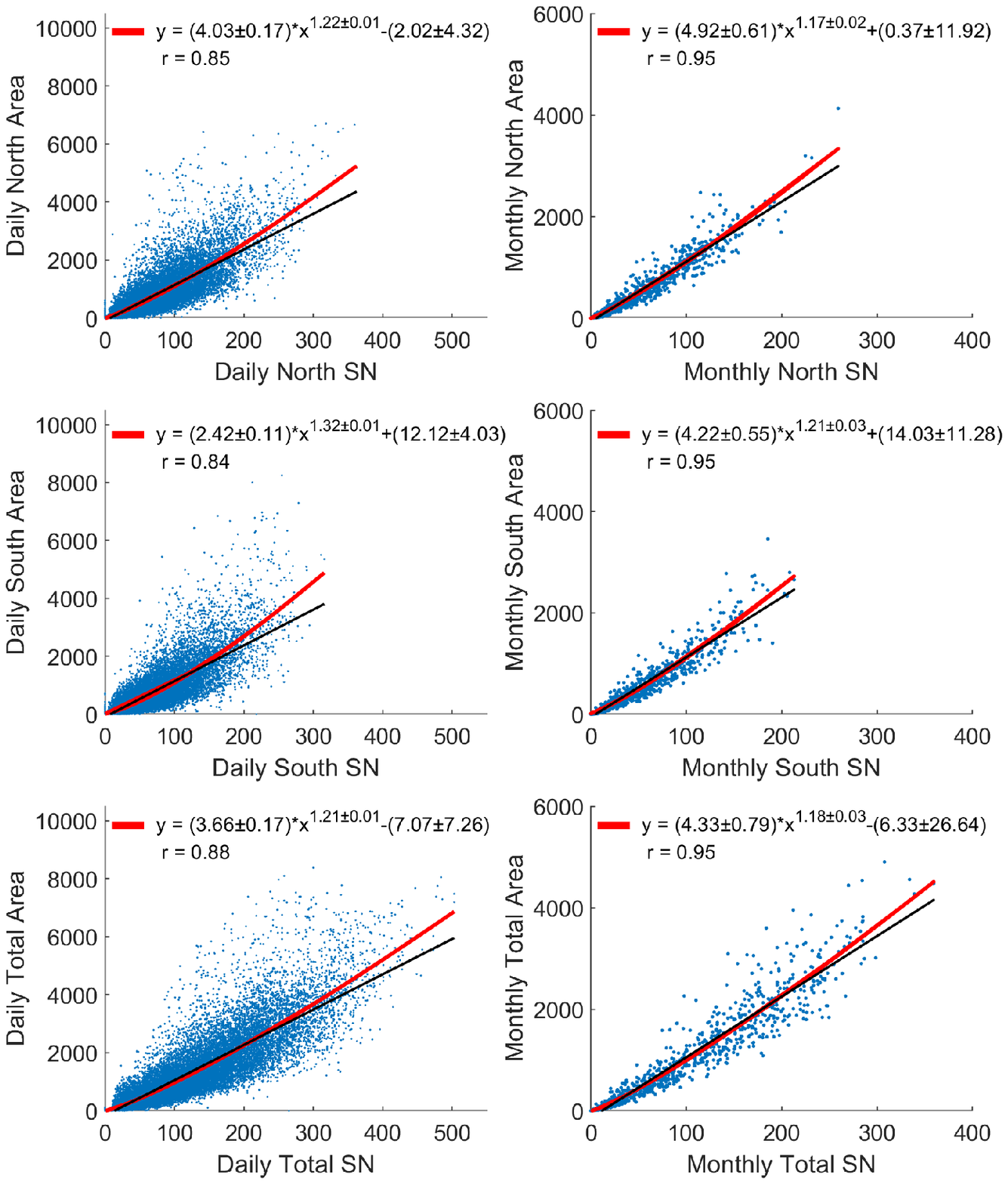}
\caption{
     Scatter plots of daily (left) and monthly mean (right) sunspot areas against sunspot numbers for the time range 1945--2016. From top to bottom: North and south HSN (from T06 and SILSO) and ISN (from SILSO).  The black line represents the linear least-squares fit, the red line the power-law fit to the data. The inset gives the power-law fit parameters and the Pearson correlation coefficient $r$.}
\label{fig:area_sn}
\end{figure}

In Figure \ref{fig:fig4} (top panels), we plot the distributions of the differences between the daily (monthly) values of the direct and reconstructed HSNs. The histograms  demonstrate that overall the data reveal no systematic differences, and the distributions are close to each other, with a mean difference of 0.5.
In the lower panels, we plot the daily (monthly) differences as a function of time. As expected, they show a correspondence to the solar cycle evolution. For the daily data, the absolute differences can be fairly large, reaching values of $>$50 around the solar cycle peaks (strongest for cycles\ 19 and 22). The differences are substantially reduced for the monthly mean data. For the daily data, the differences are characterised by an rms of 18.7. The monthly mean data give an rms of 11.5, and the smoothed monthly mean data give an rms of 4.1. 
Figure \ref{fig:area_sn}  shows the relation between the sunspot areas (from RGO) and the sunspot numbers (from T06 and SILSO) for the time range 1945--2016, together with a power-law (red line) and linear (black line) least squares fit to the data.  As one can see, the sunspot areas and numbers reveal some non-linear relationship, with the power-law indices in the range of about 1.2--1.3.  The non-linear relation between total sunspot areas and sunspot numbers has also  been recently shown for 111 years of sunspot data from the Kodaikanal Observatory \citep{Ravindra2020}, one of the stations that contributed to the RGO data. Such a non-linearity between the two indices is somewhat expected, 
as there is an intrinsic difference in weighting of solar activity between sunspot areas and sunspot numbers. There is no single number that relates the two indices, but the conversion factor may be different for low and high solar activity levels \citep[see also][]{Wilson2006}. 

However, applying the `conversion' to the relative fractions of sunspot areas in the northern and southern hemispheres, this becomes a second-order effect, and is thus smaller than for the direct area-to-sunspot relation. This is because, in general, when solar activity is high, it will be high in both hemispheres, and when it is low, it will be low in both hemispheres, which then cancels out in the relative fractions. 
In the present study, we derive the relative fractions of the daily sunspot areas in the northern and the southern hemispheres, and apply those ratios to the ISN in order to reconstruct hemispheric sunspot numbers. This requires us to use a daily conversion factor that reflects the individual level of solar activity. Comparing the scatter plots in Figure \ref{fig:corr} to those in Figure \ref{fig:area_sn}, one can see that this approach indeed largely reduces the non-linearity when reconstructing the HSN (Fig.\ \ref{fig:corr}), but still a small residual effect remains for very high activity levels. 

\begin{figure}
 \centering
\includegraphics[width=1\columnwidth]{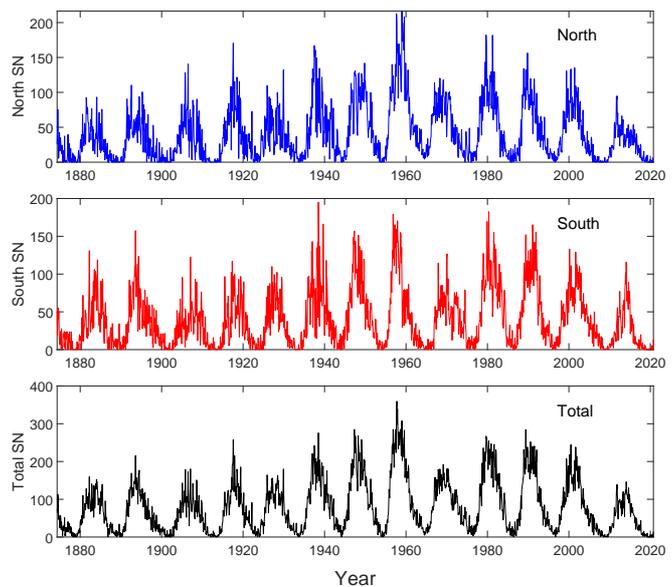}
\caption{Final series of monthly mean HSN derived for the period 1874--2020. Top: Northern hemisphere. Middle: Southern hemisphere. Bottom: Total sunspot numbers.}
\label{fig:fig5}
\end{figure}

\begin{figure}
 \centering
\includegraphics[width=1\columnwidth]{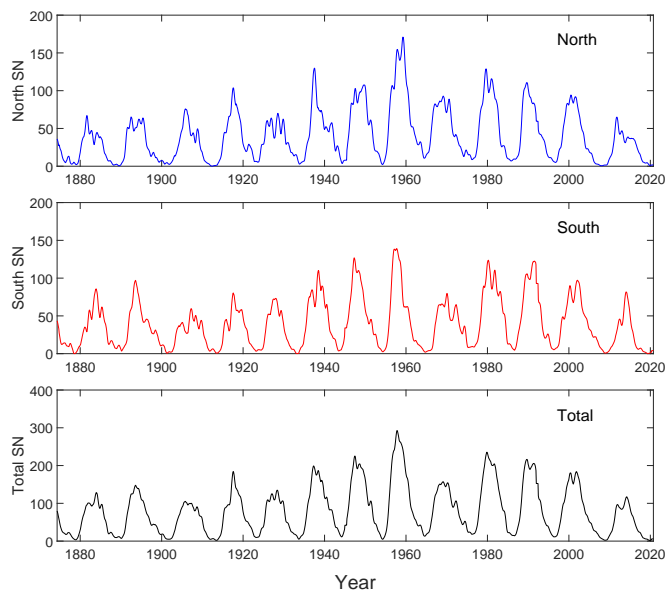}
\caption{Final series of 13-month smoothed monthly mean HSNs derived for the period 1874--2020. Top: Northern hemisphere. Middle: Southern hemisphere. Bottom: Total sunspot numbers.}
\label{fig:fig6}
\end{figure}

\section{Catalogue of hemispheric sunspot numbers 1874--2020}

Figure \ref{fig:fig5}  shows the final combined time-series of the monthly  mean HSNs from 1874--2020 together with the ISN series. Figure \ref{fig:fig6} shows the same for the smoothed monthly mean values. The final hemispheric data plotted in Figs.\ \ref{fig:fig5} and \ref{fig:fig6} are merged from the following data sets:
(i) May 1874 to December 1944: reconstructed from hemispheric sunspot areas, (ii) January 1954 to December 1991: recalibrated HSNs from T06, (iii) January 1992 to October 2020: hemispheric sunspot numbers from SILSO.
We note that this hemispheric sunspot number series can be homogeneously and continuously expanded in the future by appending the new HSN values computed by the WDC-SILSO on a monthly basis.\footnote{\url{http://sidc.be/silso/datafiles}}

\begin{figure*}
 \centering
\includegraphics[width=0.99\textwidth]{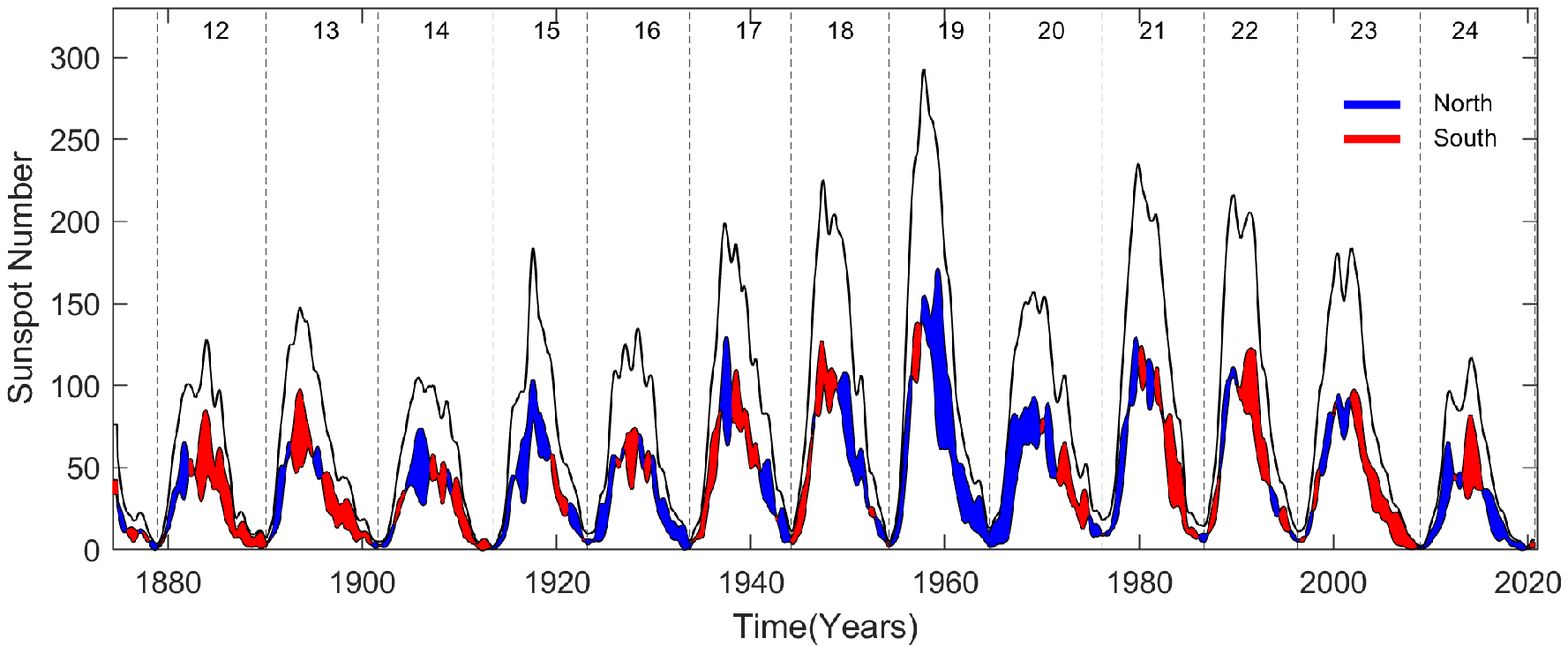}
\includegraphics[width=0.99\textwidth]{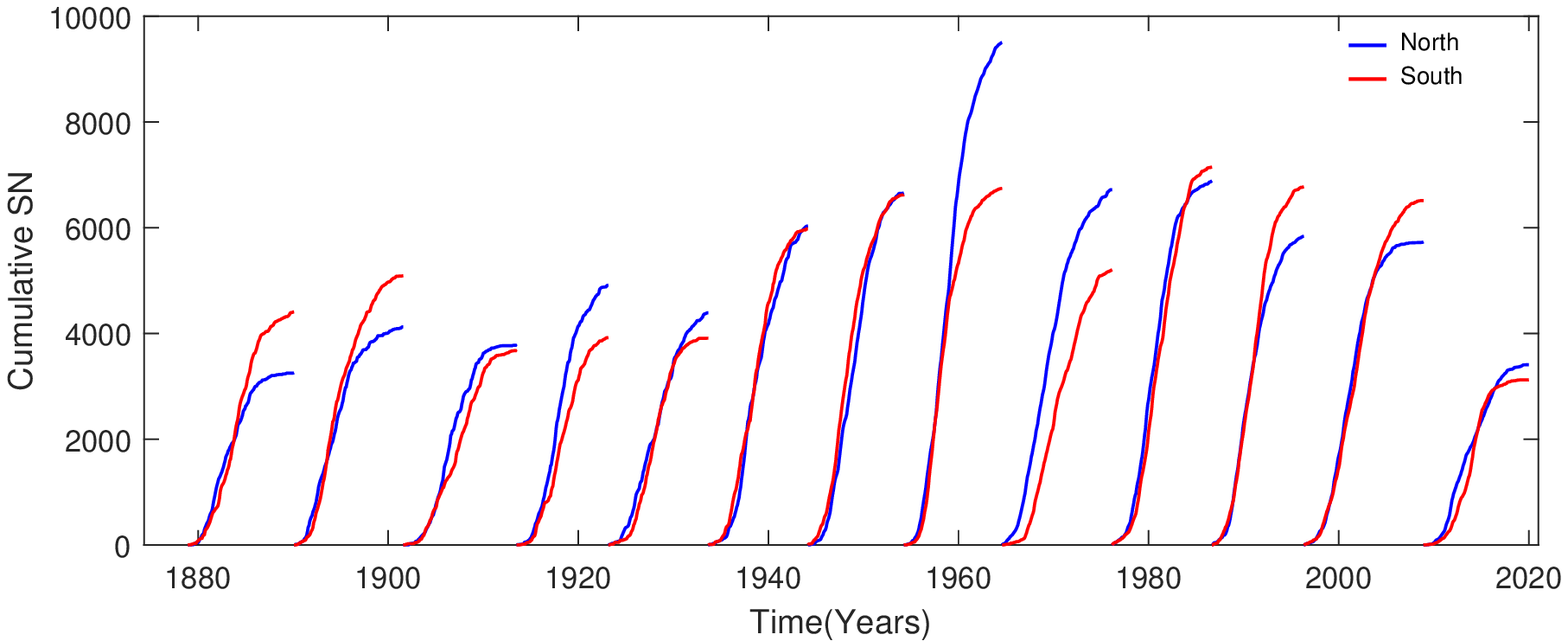}
\caption{Top: Smoothed monthly mean ISN (black) and HSN (north: blue,  south: red) from the merged data set for the time range 1874–-2020.  The shading of the difference between the north and south sunspot numbers indicates the excess between them. Dashed vertical lines are drawn at sunspot minima to delineate the individual cycles. 
Bottom:  Cumulative monthly mean HSN calculated individually for each of the cycles 12--24.}
\label{fig:hemexcess}
\end{figure*}

Accompanying this study, we provide the data compiled into an online catalogue for the time range 1874--2020. The catalogue consist of two parts: Part A compiles the daily HSN and part B compiles the monthly mean and smoothed monthly mean HSNs.
The catalogue includes the full information for the combined series derived in this study as well as for the three individual data sets used for merging. 
Catalogue A contains the daily data, and has the following entries:
\begin{itemize}
    \item 
    First column: date in the format YYYY-MM-DD.
    \item Second, third columns: combined series of daily HSNs (north, south) derived in this study (May 1874 to October 2020).
    \item Fourth, fifth columns: daily HSNs (north, south) reconstructed from sunspot areas (available for May 1874 to September 2016).
    \item Sixth, seventh columns: daily HSNs (north, south) from T06 renormalised to the recalibrated ISN (available for January 1945 to December 2004).
    \item Eighth, ninth columns: daily HSNs (north, south) from SILSO (January 1992 to October 2020)
\end{itemize}
Catalogue B contains the monthly data, and has the following entries:
\begin{itemize}
    \item First column: date in the format YYYY-MM.
    \item Second, third columns: combined series of monthly mean HSNs (north, south) derived in this study (May 1874 to October 2020).
    \item Fourth, fifth columns: combined series of 13-month smoothed monthly mean HSNs (north, south) derived in this study.
    \item Sixth, seventh columns: monthly mean HSNs (north, south) reconstructed from sunspot areas (available for May 1874 to September 2016).    
    \item Eight, ninth columns: 13-month smoothed monthly mean HSNs (north, south) reconstructed from sunspot areas. 
    \item Tenth, eleventh columns: monthly mean HSNs (north, south) from T06 renormalised to the recalibrated ISN (available for January 1945 to December 2004).
    \item Twelfth, thirteenth columns: 13-month smoothed monthly mean HSNs (north, south) from T06 renormalised to the recalibrated ISN. 
    \item Fourtennth, fifteenth columns: monthly mean HSNs (north, south) from SILSO (January 1992 to October 2020).
    \item Sixteenth, seventeenth columns: 13-month smoothed monthly mean HSNs (north, south) from SILSO. 
\end{itemize}  

The catalogue for the data presented in this study from 1874 to 2020 is available at Vizier.\footnote{\url{https://vizier.u-strasbg.fr/}} In addition, a live catalogue containing the reconstructed HSNs from this study together with the regular updated HSN will be available at SILSO.\footnote{\url{http://sidc.be/silso/extheminum}}

\section{Hemispheric asymmetry analysis and cycle growth rates}

\subsection{Hemispheric asymmetries and phase shifts}

In the top panel of Figure \ref{fig:hemexcess}, we plot the 13-month smoothed monthly mean north and south HSNs from the merged data set we created, indicating the excess by colour shadings (blue for north predominance, red for south predominance), along with the ISN data. The bottom panel shows the cumulative monthly mean HSNs, calculated individually for each of the full cycles that are covered by the data set, namely cycles\ 12--24. 

\begin{table*}
\caption{Hemispheric asymmetries for solar cycles 12--24.}          
\centering  
\small
\begin{tabular}{l c c c c c c c c c c c r r r}          
\hline\hline                        
Cycle &$t_{\rm max}$ & Max. & Max. & Max.&Rate & Rate & Rate & Cumul. & Cumul. & Predom. & Diff. & $\Delta t_{\rm N}$ & $\Delta t_{\rm S}$ & $\Delta t_{\rm {N-S}}$\\
No. & & ISN & N& S &ISN & N & S &N & S & Hem. &  [\%]&  (mon) &  (mon)&  (mon) \\ 
\hline                                   
   12 & 12/1883 & 128& 63& 84&5.3&4.4 &4.6 &3258 & 4409 & South & 35.3& $-27$ & $-1$ & $-26$ \\ 
13 & 08/1893 & 148& 64& 97&7.1 &5.6 &5.3 &4125 & 5090 & South & 23.4 & $-13$ & 0& $-13$ \\ 
14 &10/1905& 105& 74& 56 &4.3 &3.7 &3.2 &3778 & 3674 & North & \hspace{1mm} 2.8& 1& 18& $-17$  \\ 
15 &08/1917& 184& 102& 79&13.6 &6.6&7.4 &4918 & 3920 & North & 25.5&0 &1&$-1$  \\ 
16 &06/1928& 135&67&74&9.0 &5.5&4.4 &4391 & 3916 & North & 12.1 & 2&$-7$&9 \\ 
17 &05/1937&199&128&106&9.3 &8.3&5.8 &6033 & 5996 & \textit{North} & \hspace{1mm} 0.6 &2 & 14&$-12$ \\ 
18 &06/1947&225&109&125&10.2 &6.6&7.4 &6658 & 6630 & \textit{North} & \hspace{1mm} 0.4 &27 &$-1$&28  \\ 
19 &11/1957&293&168&141&13.6 &8.6&9.1 &9516 & 6741 & North & 41.2&17 &$-6$&23  \\ 
20 &02/1969&157&92&78&7.7 &5.1&4.2 &6732 & 5222 & North & 28.9&0 &12&$-12$ \\ 
21 &11/1979&235&126&122&9.0&7.0&7.1 &6889 & 7146 & South & \hspace{1mm} 3.7& $-2$& 5& $-7$  \\ 
22 &09/1989&216&111&124&12.8 &8.8&5.8 &5839 & 6780 & South & 16.1& $-1$ &22& $-23$  \\ 
23 &11/2001&184&92&98&8.0&4.6&4.7 &5731 & 6514 & South & 13.7 & $-16$ &4& $-20$ \\ 
24 &03/2014&117&64&82&7.3 &5.3&5.6 &3412 & 3126 & North & \hspace{1mm} 9.2 & $-29$ & $-1$ &$-28$ \\ 
\hline  
\hline
\end{tabular}
\label{table:table1}
\tablefoot{
We list the cycle number, the time of the cycle maximum $t_{\rm max}$ in terms of the 13-month smoothed ISN, the amplitudes of the cycle maxima in the ISN as well as north and south HSN, the peak growth rate during the cycle rise phase (in units of month$^{-1}$) for the ISN, north and south HSN,  the cumulative sunspot numbers for the northern and southern hemispheres, the predominant hemisphere, the relative difference between the cumulative sunspot numbers of the two hemispheres, the time difference between the cycle maximum (ISN) and the north HSN ($\Delta t_N$) and the south HSN ($\Delta t_S$) as well as the time difference between the HSN maxima ($\Delta t_{N-S}$). 
For cycles where the difference between the cumulated HSN is $<$1\%, we give the predominant hemisphere in italics.}
\end{table*}

The result of this analysis is summarised in Table \ref{table:table1}. We list for each cycle the total cumulated north and south sunspot numbers, the hemisphere that is predominant over the integrated cycle, and the relative difference between the cumulative numbers of the two hemispheres. As one can see, 8 out of 13 cycles reveal a difference of $>$10\%, and 5 cycles reveal a difference of $>$20\%. The mean of the relative differences over all cycles is $16.4\%$
The strongest asymmetries occurred in solar cycle 19, with a relative difference of $42\%$ between the northern and southern hemispheres in the sunspot numbers cumulated over the full cycle. This strong asymmetry for cycle 19 and also the following cycle 20 was noted by \cite{waldmeier1971} in the predominant occurrence of sunspots and faculae and also the higher coronal emission in the northern solar hemisphere. 

\begin{figure}
 \centering
\includegraphics[width=0.8\columnwidth]{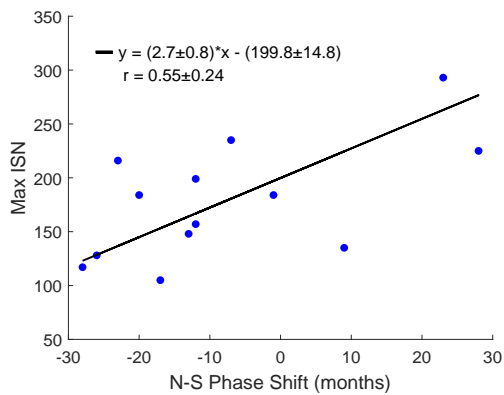}
\caption{Scatter plot of the ISN cycle maxima against the time-shift between the maxima in the north and south HSN. The black line gives the linear fit to the data.}
\label{fig:timeshift}
\end{figure}

In addition, we list in Table \ref{table:table1} the time-shift between the maxima (as determined from the 
13-month smoothed monthly mean data) of the activity in the two hemispheres, and also with respect to the ISN. The phase shifts between the north and south HSNs can be as large as 28 months. The mean absolute value of the north--south differences is 16.4 months. Considering the sign of the phase shifts for the cycles under study, there is a clear trend in that the northern hemisphere peaks earlier than the south (in 10 out of 13 cycles), with a mean value of the signed phase shifts of $-7.6$ months. This finding relates to the more global consideration of the envelopes of the hemispheric activity in \cite{McIntosh2013}, who used a 120-month moving window to cross-correlate the Greenwich hemispheric sunspot areas. These latter authors found that there exist secular trends of one hemisphere leading over the other \cite[see also][]{Zolotova2010}. In particular, they obtained that up to about 1928 the northern hemisphere is leading; between 1928 and 1968 the southern hemisphere is leading, and thereafter this switches again to the northern hemisphere. This means that for $>$70\% of the time-span under study, the northern hemisphere is leading. 

As for the maxima, we note that there  also seems to exist a weak trend in that the phase shift between the two hemispheres is correlated to the cycle amplitude (Fig. \ref{fig:timeshift}). Large shifts between the HSN where the southern hemisphere reaches its cycle maximum first are associated with stronger cycles, whereas the opposite is true (smaller amplitude cycles) when the northern hemisphere maximum is ahead. Nevertheless, we note that this trend strongly relies on the two data points with large positive time-shifts, and a larger statistical sample is needed in order to come to any firm conclusions. Also, checking the correlation of the phase shifts of the north and south HSN maxima with respect to the ISN maximum reveals that the correlation in Fig. \ref{fig:timeshift} results purely from the north HSN.

In contrast to the asymmetries in the HSN peak times, the overall HSN asymmetries (in terms of mean values for the north and south cumulated SN listed in Table \ref{table:table1}) are small (N: 5483, S: 5320), that is,\ their difference is about 3$\%$, and the mean over the HSN maxima (amplitudes) is almost identical for northern and southern hemispheres 
with a value of 97. Finally, we note that there is no obvious relation between the cumulated HSN asymmetry and the phase shifts of the maximum activity between the two hemispheres.

\subsection{Hemispheric cycle growth rates and relation to solar cycle amplitudes}

In this section, we investigate whether the separate consideration of the sunspot evolution in the two hemispheres can provide us with additional insight into the solar cycle process. \cite{cameron2008} studied, using different global indices  of solar activity (ISN, group sunspot numbers, sunspot areas, and 10.7 cm radio flux), the statistical relationship between the mean growth rate of activity in the early phase of a solar cycle and
its amplitude. For all indices, these authors found  distinct correlations, which means that stronger cycles tend to rise
faster. This can be interpreted as a dynamic variant of the \cite{waldmeier1935} rule, which relates the rise time of a cycle to its amplitude. These relations have important implications for empirical solar cycle prediction methods. 

\begin{figure}
 \centering
 \includegraphics[width=1.0\columnwidth]{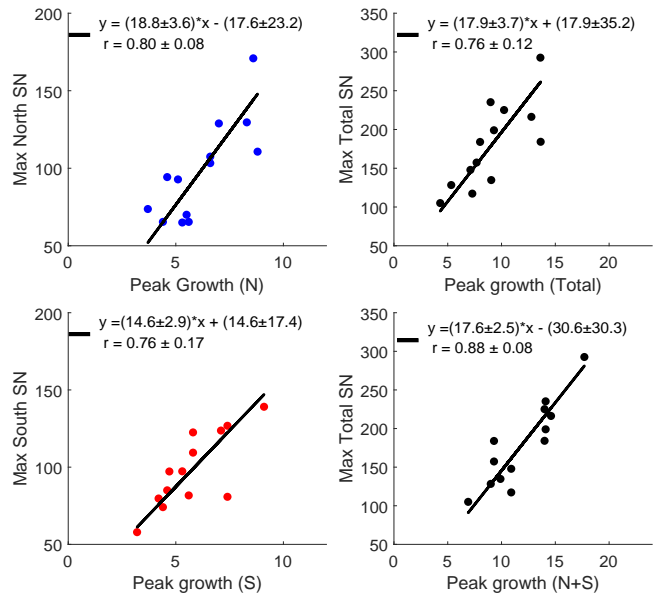}
\caption{Solar cycle amplitude against the peak growth rates for the merged hemispheric sunspot number series. Left:  
Cycle amplitude as a function of peak growth rate for north (top) and south (bottom) HSN. Top right: Cycle amplitude as a function of peak growth rate of the ISN. Bottom right: Cycle amplitude as a function of the sum of the peak growth rates determined separately for the north and the south HSN. }
\label{fig:growth_merged}
\end{figure}

Here, we make a similar type of analysis, but with two main changes. First, we apply the analysis not only to the ISN but also separately to the HSN in order to test whether these may lead to improved relations. Second, we do not study the mean growth rate over a fixed amplitude range of the solar cycle but instead we derive the peak growth rate in the cycle's rise phase from the time derivative of the smoothed sunspot number series. This is because the fixed range as used in \cite{cameron2008} implies, for smaller cycles, that the limit is only reached shortly before or even at the cycle maximum, which means that for prediction purposes the lead time would become zero. Also, we find that the peak growth rates give higher correlations than the mean growth rates.

In Figure \ref{fig:growth_merged}, we show the scatter plots of the solar cycle amplitudes against the peak growth rates derived from the 13-month smoothed sunspot numbers. These numbers are also listed in Table \ref{table:table1}. The black lines show the linear fits to the data that were derived using the least-squares bisector regression method in order to symmetrically consider the variables on the $x$- and $y$-axes \citep{Isobe1990}. The inset gives the fit parameters as well as the Pearson correlation coefficient. Error ranges on the correlation coefficients were determined via the bootstrapping method using 10,000 realisations for each of the relations \citep{Efron1993}.

We study the cycle amplitudes for the HSNs of our merged data set against the corresponding hemispheric peak growth rates. For comparison, the same is done for the ISN data. In addition, we also evaluate the ISN cycle amplitude as a function of the sum of the peak growth rates determined separately for the north and south HSN (bottom right panel in Fig. \ref{fig:growth_merged}). 
As the peaks in the time derivatives of the northern and southern sunspot numbers in general do not occur at the same time, this linear combination of the two peaks is different from calculating the peak growth rate of the linear combination of the hemispheric sunspot times series (which is equal to the ISN series).
We note, for both panels on the right hand side, we plot on the $y$-axis the solar cycle amplitude as derived from the ISN. 
As one can see from these scatter plots and insets, the Pearson correlation coefficient has increased from $r=0.76\pm 0.09$ to $r=0.88\pm 0.08$ when deriving the peak growth rate separately for the two hemispheres and combining them afterwards, instead of deriving the peak growth rate from the total sunspot number series. 
These findings imply that some information gets lost when we consider the total sunspot number series and its dynamics instead of the dynamics of the two hemispheres individually.

\begin{figure}
 \centering
\includegraphics[width=1.0\columnwidth]{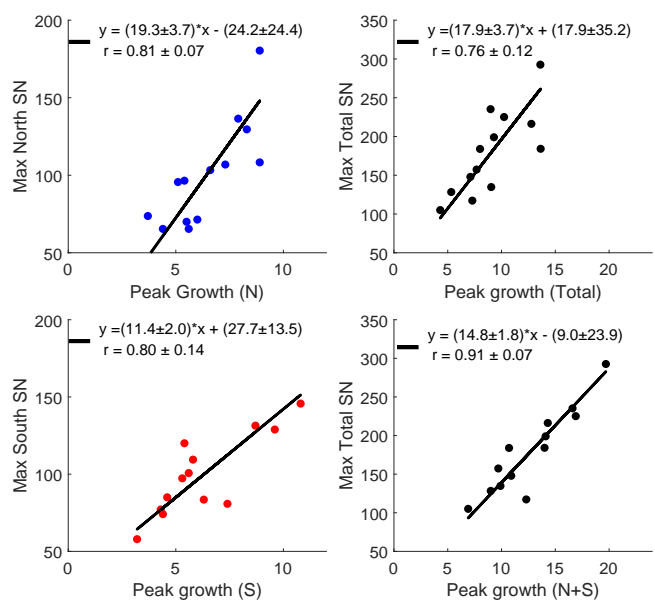}
\caption{Solar cycle amplitude against the peak growth rates for the purely area-based hemispheric sunspot number series. Left:  
Cycle amplitude as a function of peak growth rate for north (top) and south (bottom) HSN. Top right: Cycle amplitude as a function of peak growth rate of the ISN. Bottom right: Cycle amplitude as a function of the sum of the peak growth rates determined separately for the north and south HSN. 
}
\label{fig:growth_areas}
\end{figure}

Interestingly, this phenomenon shows up even stronger when instead of the merged hemispheric sunspot series, we use the HSN purely derived from the hemispheric sunspot areas, where the correlation increases from $r=0.76\pm 0.12$ to $r=0.91\pm 0.07$ (Figure \ref{fig:growth_areas}). 
What is also noticeable from these plots is that in both cases, the dependence of the cycle amplitude from the peak growth rate is steeper for the northern than the southern HSN, with slopes of 19 versus\ 15 (Fig.\ \ref{fig:growth_merged}) and 19 versus\ 11 (Fig.\ \ref{fig:growth_areas}). However, we note that this effect is strongly related to cycle\ 19 and its high northern HSN growth rate, which stands out even more for the reconstructed HSN.
This is an interesting finding and is most probably related to the remaining non-linearity between direct and reconstructed HSNs (Fig. \ref{fig:corr}), which is expected to be most significant when both the asymmetry and the activity are high. These conditions are indeed most strongly met in cycle\ 19. Overall, both data sets give similar results.

We also note that the peak growth rates show considerably closer correlations than the mean growth rates in the cycle's rise phase. However, in this case the separate consideration of the HSN  also improves the results compared to the ISN growth rates (though the overall level is lower).  \cite{cameron2008} studied the mean growth rate  of the ISN ---derived
within a fixed range of ISN (Version 1.0) between 30 and 50 in the cycle's rise
phase--- against the cycle amplitude. 
Here we transform this range from the ISN Version 1.0 to 2.0 by multiplying it by a mean scaling factor of 1.4, resulting in a range of  42--70. For the HSN, we adopt a corresponding range of 21--35. The scatter plots of the mean growth rates against the cycle amplitude are shown in Fig.\ \ref{fig:growth_mean}. As one can see, the correlation coefficients are considerably lower than those for the peak growth
rates (Figure \ref{fig:growth_areas}), with  $r=0.53\pm 0.24$ and $r=0.62\pm 0.16$ for the ISN and combined HSN relations, respectively. This implies that the peak growth rate encodes more information about the solar cycle amplitude than any average growth rate.

To investigate a) why the sunspot numbers derived from the hemispheric areas give a higher correlation than the merged series and b) why the slopes derived  for the Northern and Southern hemispheres are different, we show in the Appendix the same plots as in Figs. \ref{fig:growth_merged} and \ref{fig:growth_areas}, but separately for the time ranges where in the merged data series the HSNs are derived from the sunspot areas (cycles 12--17) and from the T06 and SILSO data (cycles 18--24). Figure\ \ref{fig:growth_merged_part2} shows the scatter plots for cycles 18--24 using the HSNs from the merged series. Figure\ \ref{fig:growth_reconstructed_part2} shows the scatter plot for the same cycles but for the HSNs reconstructed from the area data. Figure\ \ref{fig:growth_reconstructed_part1} shows the scatter plot for cycles 12--17, where we only have HSNs reconstructed from the area data. These plots show that the dependence between cycle peak amplitude and growth rate is always steeper for the northern hemisphere than for the southern hemisphere, and also that this effect is stronger for the area-based data-series
than for the merged series. These figures also show that, for each subset, the correlation with the cycle amplitude is higher when considering the sum of the HSN peak growth rates than that of the ISN. The most specific subrange is the one covering solar cycles 12--17. Here we observe the largest difference in the slopes between north and south,  and summing the peaks of the derivatives of the north and south hemispheric sunspot series leads to a very high correlation with the solar cycle amplitude of $r=0.99 \pm 0.01$. 
Combining the data sets of the different subranges, which are characterised by different slopes, leads to a lower overall correlation coefficient for the full range, and has a bigger effect on the merged data series.

\section{Discussion and conclusions}

In this study, we created a continuous series of daily, monthly mean, and smoothed monthly mean HSNs for the period 1874--2020, fully covering solar cycles 12--24 (Figs. \ref{fig:fig5} and \ref{fig:fig6}). The data are compiled in a data catalogue that accompanies the paper.\footnote{\url{https://vizier.u-strasbg.fr/}} In the future, this data series will be seamlessly expanded with the HSNs that are regularly provided by SILSO.\footnote{\url{http://sidc.be/silso/extheminum}} The hemispheric sunspot number series is combined from three data sources: Starting with the year 1945, it contains the HSN provided by T06 and SILSO. The reconstructions back in time to 1874 are based on the 
daily measurements of hemispheric sunspot areas from GRO and NOAA. Validating our reconstructions with the direct hemispheric sunspot data for the time range 1945--2016 reveals very good agreement (see Fig. \ref{fig:corr}), with correlation coefficients of $r=0.94$ for the daily data and $r=0.97$ for the monthly mean data. The power-law fit indicates a relation that is close to linear, except for the highest activity levels where some small non-linearity effects show up. 

For the cumulative hemispheric sunspot number asymmetries (Fig. \ref{fig:hemexcess}, bottom panel) we obtain a mean value of 16\% over cycles 12--24. The strongest asymmetry occurs in cycle\ 19 (which is the highest cycle in the period under study), where the northern hemisphere shows a  predominance as high as 42\% cumulated over the solar cycle. 
The phase shift between the peak of the solar activity in the northern and southern hemispheres may be as large as 28 months, with a mean value over all cycles under study of 16.4 months. Interestingly, there exists a distinct hemispheric asymmetry in that, in 10 out of 13 cases, the northern hemisphere reaches its cycle maximum earlier, with  a mean of the signed phase shift over all cycles under study of $-7.6$ months. Also, there are indications that this hemispheric phase shift is correlated to the cycle amplitude. 
Finally, we demonstrate that the dynamic Waldmeier rule, which relates the growth rate of a solar cycle to its amplitude, works better when the two hemispheres are considered separately. To this aim, we determined the peak of the time derivative of the HSN during the cycle's rise phase, and combined them afterwards. This yielded higher correlations with the solar cycle amplitude than calculating the peak of the time derivative of the total sunspot numbers, with  correlation coefficients of $r \approx 0.9 \pm 0.08$ compared to $r = 0.76\pm 0.12$, respectively, for the total sunspot numbers (Figs.\ \ref{fig:growth_merged} and \ref{fig:growth_areas}). 

It has been shown for various solar activity indices related to surface magnetism that there exist significant asymmetries and phase shifts in the evolution of the two hemispheres  \citep[e.g.][]{newton1955,waldmeier1971,garcia1990,antonucci1990,temmer2002, temmer2006,durrant2003,joshi2004,knaack2005,norton2010,McIntosh2013,deng2016,chowdhury2019,roy2020}. In recent years, north--south asymmetries have also been identified in  Joy's law \citep{McClintock2013} as well as in east--west zonal flows and north--south meridional flows \citep{Zhao2013,Komm2014}, which play a fundamental role in the magnetic flux transport inside the Sun. These findings suggest that the solar cycle evolution of the two hemispheres is partly decoupled, and that these asymmetries are a result of the underlying solar dynamo  \cite[e.g.][]{sokoloff1994,ossendrijver1996,tobias1997,norton2014,schuessler2018}. This also has implications for solar cycle prediction methods.

The different types of solar cycle prediction methods either rely on empirical relations or are physics-based  \cite[for reviews see,][]{Pesnell2012,petrovay2020}. Long and continuous sunspot number records are needed for predictions of solar activity on different timescales, covering short, medium, and long-term horizons (see the definitions in \cite{Petrova2021}). Long-term prediction methods mainly focus on forecasting the solar cycle amplitude, and broadly use sunspot numbers in combination with other solar and geomagnetic activity indices. A large number of methods are based on finding a particular sunspot number precursor in the past, which can serve as an indicator for the amplitude of the upcoming solar cycle \citep[e.g.][]{waldmeier1935,
Ramaswamy1977, Macpherson1995, Conway1998, Sello2001, LantosSkewness2006, cameron2008, 
PodladchikovaLefebvreLinden2008,podladchikova2017,
Aguirre2008, Braja2009,Podladchikova2011, Kakad2020}.
Other types of precursor methods are based on the polar magnetic fields \citep[e.g.][]{SchattenDynamo1978, Schatten1996, SchattenSofia1987, Svalgaard2005, Schatten2005, WangSheeley2009, MunozJaramillo2012} and 
geomagnetic activity indices  \citep[e.g.][]{OhlOhl1979, Feynman1982, GonzalezSchatten1988, Thompson1993, WilsonHathawayReichmann1998}. In the recent decades, also physics-based prediction methods were developed based on flux transport and dynamo models \citep[e.g.][]{Nandy2002, DikpatiGilman2006, CameronSchlussler2007, Choudhuri2007, Henney2012, CameronSchlussler2015}. 

Short- and medium-term predictions of solar activity are usually performed for the ongoing cycle development with lead times from days to months. SILSO regularly provides updated predictions of sunspot numbers from 1 to 12 months ahead using the `standard' method \citep{Waldmeier1968}, the `combined' method \citep{Denkmayr1997, Hanslmeier1999}, the McNish–Lincoln method \citep{McNish1949}, as well as  improvements of these prediction methods using a Kalman filter \citep{Podladchikova2012}. All these methods are based on the ISN series.

The concluding point we want to make here is that the catalogue of hemispheric sunspot number created in this study provides an extended record of the hemispheric activity evolution over 12 solar cycles, and will be homogeneously extended with future HSNs provided by SILSO. Furthermore, we demonstrate that empirical relations between the solar cycle growth rate and its amplitude show higher correlations when considering the HSN. These findings provide the foundation for enhanced empirical solar cycle prediction methods based on the HSN to capture the hemisphere's individual evolution and dynamics.

\section*{Appendix A: Additional Figures}

\begin{appendix}

In this Appendix A, we show more investigations on the relation between the total and hemispheric solar cycle growth rates with the solar cycle amplitudes. Figure\ \ref{fig:growth_mean} shows the scatter plots for the mean cycle growth rates.
Figures \ref{fig:growth_merged_part2}-- \ref{fig:growth_reconstructed_part1} show the relations of the peak growth rates
 and the cycle amplitudes separately for the different time-spans, for which we have direct measurements of HSN available (cycles 18--24; Figs.\ \ref{fig:growth_merged_part2}, \ref{fig:growth_reconstructed_part2})  
and where we have only hemispheric area measurements available (cycles 12--17; Fig. \ref{fig:growth_reconstructed_part1}).

\renewcommand{\theequation}{A\arabic{equation}}
\setcounter{equation}{0}
\renewcommand{\thefigure}{A\arabic{figure}}
\setcounter{figure}{0}

\begin{figure}
 \centering
\includegraphics[width=1.0\columnwidth]{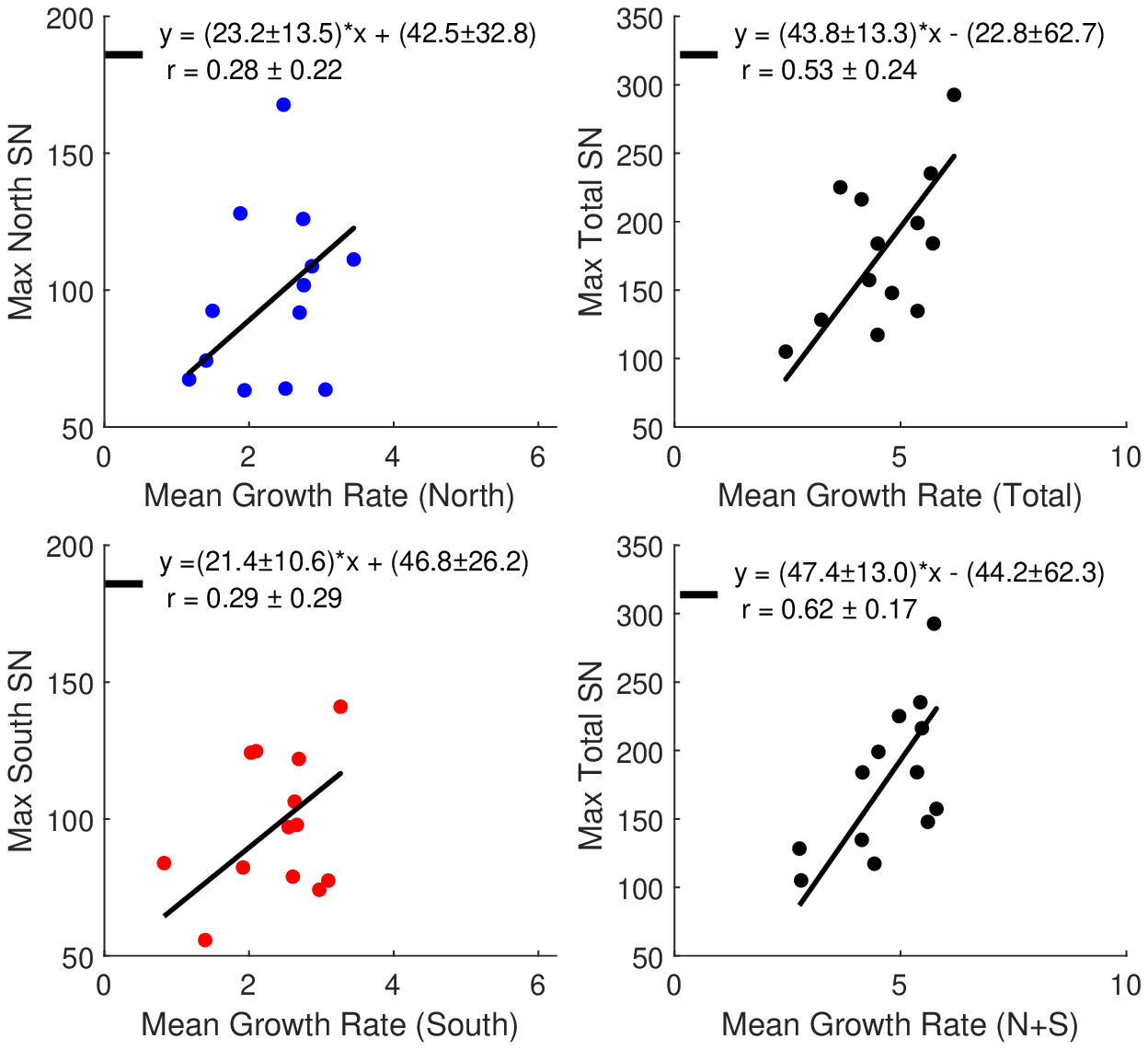}
\caption{Same as Fig. \ref{fig:growth_merged} (merged hemispheric sunspot number series) but for mean growth rates.}
\label{fig:growth_mean}
\end{figure}

\begin{figure}
 \centering
\includegraphics[width=1.0\columnwidth]{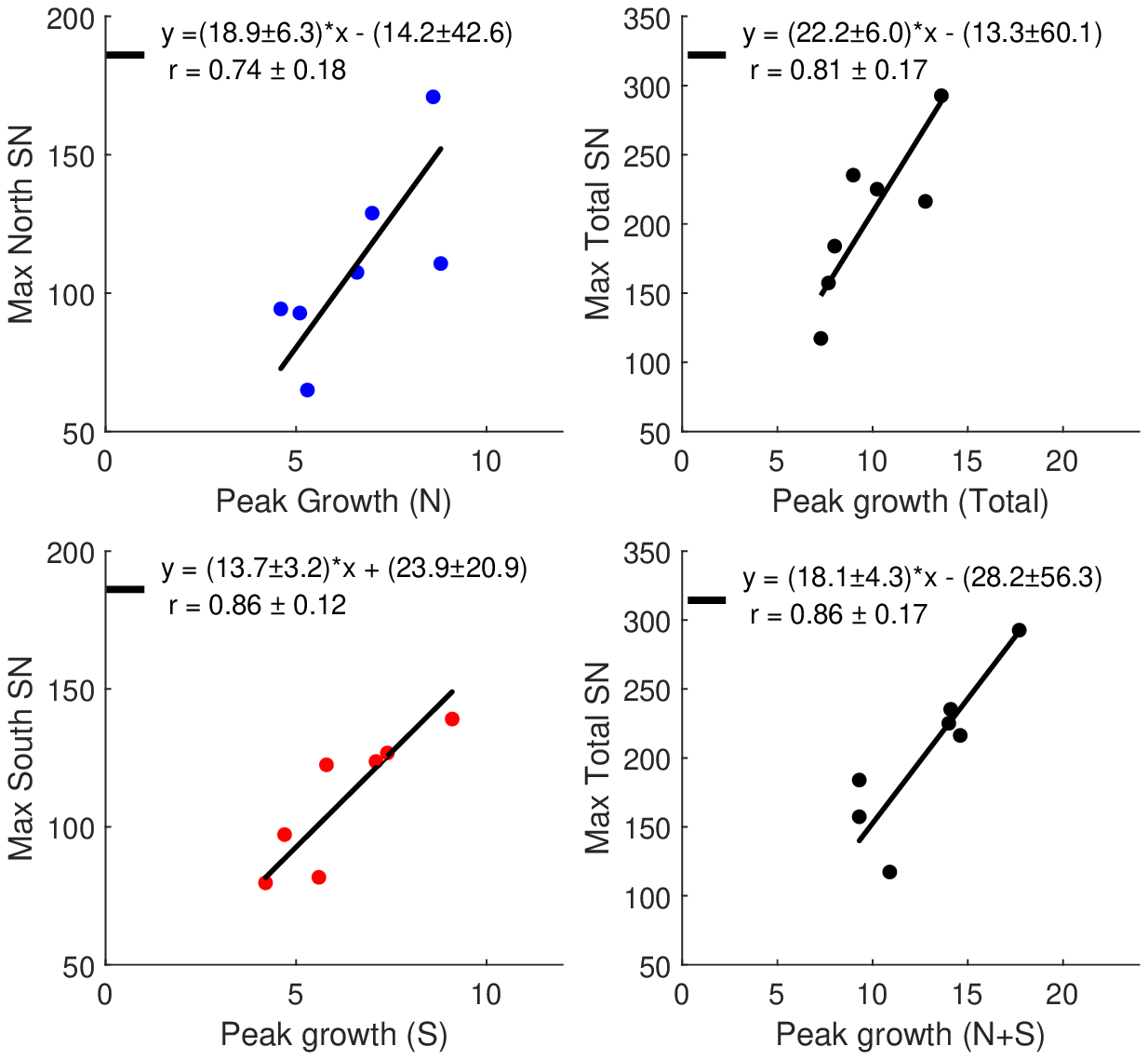}
\caption{Same as Fig. \ref{fig:growth_merged} (merged hemispheric sunspot number series) but solely for cycles 18--24.}
\label{fig:growth_merged_part2}
\end{figure}

\begin{figure}
 \centering
\includegraphics[width=1.0\columnwidth]{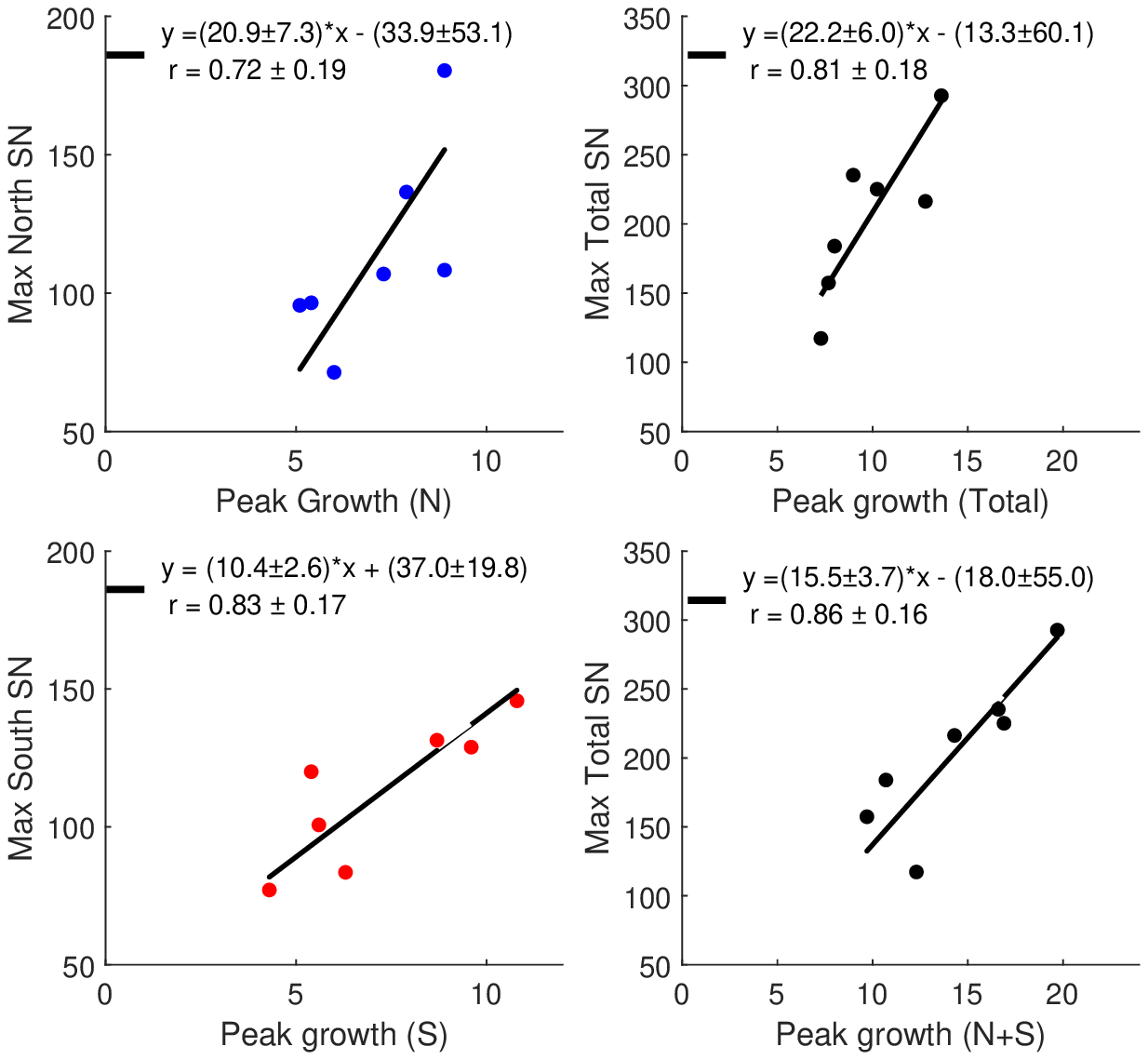}
\caption{
Same as Fig. \ref{fig:growth_areas} (purely area-based hemispheric sunspot number series) but solely for cycles 18--24.}
\label{fig:growth_reconstructed_part2}
\end{figure}

\begin{figure}
 \centering
\includegraphics[width=1.0\columnwidth]{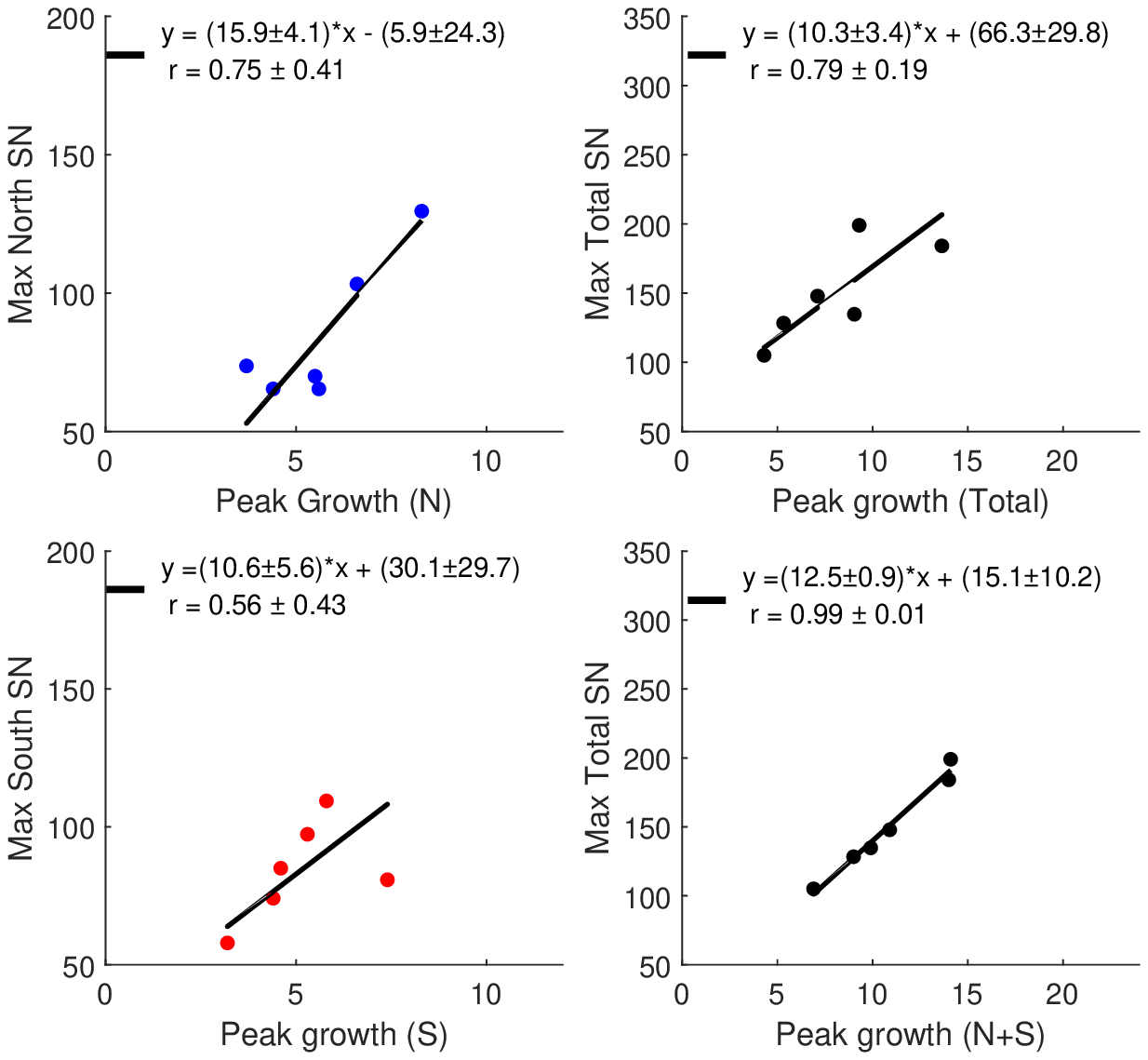}
\caption{Same as Fig. \ref{fig:growth_areas} (purely area-based hemispheric sunspot number series) but solely for cycles 12--17.}
\label{fig:growth_reconstructed_part1}
\end{figure}

\end{appendix}
\begin{acknowledgements}
      The World Data Center SILSO, which produces the international sunspot number used in this study, is supported by Belgian Solar-Terrestrial Center of Excellence (STCE, \url{ http://www.stce.be}) funded by the Belgian Science Policy Office (BelSPo). This research has received financial support from the European Union’s Horizon 2020 research and innovation program under grant agreement No. 824135 (SOLARNET).
\end{acknowledgements}

\bibliographystyle{aa}

\end{document}